# Electrical homogeneity of large-area chemical vapor deposited multilayer hexagonal boron nitride sheets

*Fei Hui,\*,§,‡ Wenjing Fang,§,‡ Wei Sun Leong, §,& Tewa Kpulun,ξ Haozhe Wang,§ Hui Ying Yang,& Marco A. Villena,\* Gary Harris,ξ Jing Kong,§ Mario Lanza\**

\*Institute of Functional Nano & Soft Materials, Collaborative Innovation Center of Suzhou Nano Science and Technology, Soochow University, Suzhou, 215123, China.

§Department of Electrical Engineering and Computer Science, Massachusetts Institute of Technology, Cambridge, MA 02139, USA.

&Pillar of Engineering Product Development, Singapore University of Technology and Design, 8 Somapah Road, Singapore 487372, Singapore

ξDepartment of Electrical and Computer Engineering, Howard University, Washington DC, 20059, USA.

δDepartment of Materials Science and Engineering, Stanford University, CA 94305, USA

KEYWORDS: hexagonal boron nitride, chemical vapor deposition, electrical homogeneity, conductive AFM, polycrystalline.

ABSTRACT

Hexagonal boron nitride (*h*-BN) is a two dimensional (2D) layered insulator with superior dielectric performance that offers excellent interaction with other 2D materials (*e.g.* graphene, $MoS_2$). Large-area *h*-BN can be readily grown on metallic substrates via chemical vapor deposition (CVD), but the impact of local inhomogeneities on the electrical properties of the *h*-BN and their effect in electronic devices is unknown. Here it is shown that the electrical properties of *h*-BN stacks grown on polycrystalline Pt vary a lot depending on the crystalline orientation of the Pt grain, but within the same grain the electrical properties of the *h*-BN are very homogeneous. The reason is that the thickness of the CVD-grown *h*-BN stack is different on each Pt grain. Conductive atomic force microscopy (CAFM) maps show that the tunneling current across the *h*-BN stack fluctuates up to 3 orders of magnitude from one Pt grain to another. However, probe station experiments reveal that the variability of electronic devices fabricated within the same Pt grain is surprisingly small. As cutting-edge electronic devices are ultra-scaled, and as the size of the metallic substrate grains can easily exceed 100 μm (in diameter), CVD-grown *h*-BN stacks may be useful to fabricate electronic devices with low variability.

MAIN TEXT

The integration of two dimensional (2D) layered materials in electronic devices is a very successful strategy to enhance their intrinsic performance, as well as to provide additional functionalities such as flexibility and transparency.[1] Conductive and semiconducting layered materials (*e.g.* graphene, $MoS_2$) have been already integrated in many types of devices, including field effect transistors (FETs),[2] supercapacitors and sensors.[3] In all these prototypes, the 2D layered materials need to interact with dielectrics (insulators) in order to provide electrical fields and capacitance effects. Unfortunately, the interaction between 2D materials (*e.g.* graphene, $MoS_2$) and traditional dielectrics (*e.g.* $SiO_2$, $HfO_2$, $Al_2O_3$) is very problematic, due to defective bonding and high scattering at the interface.[4] Hexagonal boron nitride (*h*-BN) is a layered insulator (direct band gap ~ 5.9 eV),[5] in which boron and nitrogen atoms arrange in a $sp^2$ hexagonal structure by covalent bonding, whereas the layers stick to each other by van der Waals attraction. Given its high in-plane mechanical strength (500 N/m),[6] large thermal conductivity (600 $Wm^{-1}K^{-1}$),[7] and high chemical stability (up to 1500 ºC in air),[8] *h*-BN has attracted much attention for a wide range of potential applications. For example, given its ultra-flat and surface free of dangling bonds *h*-BN substrates can increase the mobility of graphene-based FETs up to ~140,000 $cm^2V^{-1}s^{-1}$ [9] (on $SiO_2$ substrates it is lower, 15,000 $cm^2V^{-1}s^{-1}$). When used as dielectric, *h*-BN has shown enhanced reliability (compared to $HfO_2$),[10] characteristic layer-by-layer dielectric breakdown process,[11] and resistive switching.[12]

The first work on 2D *h*-BN (dating from 2005) isolated multilayer nanosheets (area <100 μm$^2$) via mechanical exfoliation.[13] As this technique is non-scalable, developing other methodologies for *h*-BN fabrication is of utmost importance.[9] Liquid-phase exfoliation, molecular beam epitaxy and physical vapor deposition have been also suggested for *h*-BN preparation,[11] but these approaches lead to abundant local defects and dangling bonds, which reduce the overall performance of the *h*-BN. Chemical vapor deposition (CVD) is an attractive approach to synthesize large-area *h*-BN stacks with low density of defects, in which the size of the samples is only limited by the size of furnace tube. Monolayer *h*-BN was firstly synthesized via CVD in ultra high vacuum (UHV) systems on single-crystal transition metals, such as Pt (111)[14] and Ni (111),[15] but the complexity of the UHV system hindered its widespread. Later reports presented the CVD synthesis of monolayer sheets and few-layers-thick stacks of *h*-BN on metallic foils (*e.g.* Cu, Fe, Pt) at atmospheric pressure (APCVD) and low pressure (LPCVD).[16-19] It should be highlighted that, due to the high temperatures required for the CVD growth of *h*-BN (>800ºC), the metallic substrates become polycrystalline. Ref.[19] reported that the thickness of the *h*-BN stack grown by CVD on polycrystalline Ni depends on the size of the Ni crystal underneath, and it was also suggested that *h*-BN grows faster on the surface of Ni (100) than Ni (111), due to the different catalytic reaction activities.[20] Similar observations have been recently reported for *h*-BN stacks grown via CVD on polycrystalline Pt substrates.[21] However, the impact of these thickness fluctuations on the local electrical properties of the *h*-BN stacks and their effect on the performance of electronic devices is still unknown. This information is essential to understand and control the device-to-device variability, which has been reported as one of the major problems of ultra-scaled technologies.[22]

In this work, the electrical homogeneity of continuous, large-area and high-quality *h*-BN stacks (grown by LPCVD on Pt substrates) is investigated via conductive atomic force microscopy (CAFM) and probe station. We find that thicker *h*-BN preferably grows on Pt grains with (101) crystallographic orientation. The excellent topographic-current correlation observed in CAFM maps indicates that the tunneling current across the *h*-BN is homogeneous within each Pt grain, but very different from grain to grain. Device level tests revealed that the variability of the devices fabricated within the same Pt grain is very small, and that the properties of devices grown on different Pt grains are strongly different to each other. The results here presented provide new insights on the electrical homogeneity of large-area *h*-BN stacks, and contribute to understand the variability of *h*-BN based electronic devices.

Figure 1a shows the schematic of the LPCVD process for *h*-BN growth. A 1 cm × 2.5 cm Pt substrate was cleaned via thermal annealing (see experimental section) and introduced in the center of the CVD tube (see Figure 1a). We use Pt as substrate because, despite being more expensive than Cu, Ni or Fe, the quality of the *h*-BN grown on Pt is higher (*i.e.* it holds a better layer structure with less randomly oriented crystallities).[23] Liquid-phase borazine precursor was kept in a commercial cold container at 3 ºC to avoid it's self-decomposition, and a cold trap filled with liquid nitrogen was used to prevent the damage of the pump. The borazine molecules (0.1 sccm) were delivered to the Pt substrate on $H_2$ carrier gas (70 sccm) at 950 ºC, which produced their absorption and decomposition on the surface of the Pt substrate, and the self-mediated growth of *h*-BN. After the LPCVD growth, the *h*-BN stack was transferred onto 300-nm-$SiO_2$/Si for Raman spectroscopy and optical microscopy inspection, and on metallic grids for transmission electron microscopy (TEM) characterization. The transfer of the *h*-BN was carried out following the bubbling approach based on water electrolysis (see Figure 1b and the experimental section).[23] This method is beneficial because it allows recycling the Pt foil for unlimited times (*i.e.* it is cost-effective).

Figure 2a and 2b show the optical image and electron backscatter diffraction (EBSD) map of a Pt substrate after the thermal annealing (right before *h*-BN growth). Different Pt grains and crystallographic orientations can be distinguished. A long marker at the bottom right part of the image was made with a razor blade to identify this location in subsequent analyses. After that, multilayer *h*-BN was grown directly on the surface of the Pt substrate and transferred onto flat 300-nm-$SiO_2$/Si. The optical image of the *h*-BN/300-nm-$SiO_2$/Si (Figure 2c) reveals regions with different colors (dark-green to light-green) perfectly matching the shapes of the Pt grains before *h*-BN growth (Figures 2a and 2b). Figure 2c also proves that the *h*-BN stack is continuous, as well as the non-destructive nature of the bubbling transfer method. Interestingly, the *h*-BN stacks grown on Pt grains with crystallographic orientations near (101) show light-green colors in the optical microscope image (see Figure 2c), which has been attributed to a larger thickness.[19,21] This hypothesis has been verified by collecting Raman spectra at different locations on the surface of the *h*-BN/300-nm-SiO$_2$/Si sample (see Figure 2d). The dark-green grains in the optical microscope image (Figure 2c) always showed an E$_{2g}$ peak near 1370 cm$^{-1}$, which corresponds to monolayer *h*-BN; on the contrary at the light-green grains the E$_{2g}$ peak shifted towards ~1366 cm$^{-1}$, which is the characteristic value of bulk *h*-BN. High magnification TEM images (Figure 2e) present the definitive corroboration of grain-dependent thickness in the *h*-BN stack. The thicknesses observed via TEM at multiple locations of the sample (Figure 2e) always ranged between 1-2 and 10-13 layers, meaning that the dark and light regions in Figure 2c should correspond to thicknesses of 1-2 and 10-13 layers, respectively. It is worth noting that, for all thicknesses, the TEM images reveal defect-free layered structure. The electron energy loss spectroscopy (EELS) spectrum (Figure 2f) shows two edges at around 180 and 390 eV, which correspond to the characteristic k-shell ionization edges of boron and nitrogen. This verifies that the stoichiometric ratio of boron and nitrogen is 1:1, which is characteristic of layered *h*-BN with hexagonal lattice.[6]

The surface roughness of the *h*-BN grown on Pt grains with different orientations has been analyzed via atomic force microscopy (AFM). Figure S1 shows the topographic AFM maps measured on the surface of the 300-nm-SiO$_2$/Si sample shown in Figure 2c; the root mean square (RMS) roughness of each image (calculated via AFM software) is also displayed. As it can be observed, the flattest surface is detected for the *h*-BN that was grown on Pt (111), and the roughest corresponds to the *h*-BN that was grown on Pt (101). As the surface roughness of 2D materials increase with their thickness,[24] Figure S1 further supports that the thinnest *h*-BN stack grows on Pt (111), and that the thickest grows on Pt (101). Moreover, multilayer islands have been detected on the surface of the *h*-BN grown on Pt (001) and Pt (101), as shown in **Figure S2**, further suggesting the growth of thicker *h*-BN stacks. Alternative methods to evaluate the thickness of the h-BN stack are low energy electron microscopy (LEEM) and low energy electron diffraction (LEED),[25] which may be used in future works.

The electrical performance of as-grown *h*-BN/Pt (without transfer) has been tested via CAFM. Figures 3a and 3b show the simultaneously collected topographic and current maps (respectively) measured on the surface of the *h*-BN stack when applying a bias of -2V to the Pt substrate (CAFM tip grounded). In order to analyze the electrical properties of *h*-BN grown on different Pt grains, we used a large (80 μm × 80 μm) scan size and strategically selected an area of the sample containing several Pt grains. The different Pt grains detected have been named with letters from A to G. Despite the *h*-BN cannot be detected in the topographic map (it is very flat compared to the Pt surface, see Figure S1), its effect can be clearly seen in the current map (Figure 3b), which reveals sharp conductivity changes from grain to grain. The different currents collected on each Pt grain must be related to the presence of *h*-BN stacks with different thicknesses because the underlying Pt grains, despite having different crystal orientations, hold similar conductivities.[26] It should be highlighted that the *h*-BN adhesion to the Pt surface is always by van der Waals forces for any Pt crystalline orientation, which means that the electronic coupling between the h-BN and Pt from one grain to another does not change. Therefore, h-BN/Pt electronic coupling is not one factor producing the conductivity changes from one grain to another observed in Figure 3b. The fact that the conductivity changes have been detected in the same image discards tip wearing from one grain to the other, and the perfect topography-current correlation undoubtedly demonstrate that *h*-BN grown on different grains hold different conductivities (due to their different thicknesses, as shown in Figure 2). It should be highlighted that, sporadically, the regions close to the Pt grain boundaries have shown higher currents (see for example the grain boundaries between grains D-G and F-G). The explanation for this observation is as follows: when the metallic substrate is exposed to large temperatures it becomes polycrystalline; the grain boundaries of the metal substrate are rough and may contain asperities; these topographic accidents alter the h-BN growth, and at that location the h-BN may be thinner or even cracked, displaying large currents in the current maps. Cross-sections have been collected (offline) at all the grains of the current image (Figure 3b) using the AFM software (NanoScope Analysis) and assembled one after the other (using OriginPro 8 software). The result is displayed in Figure 3c. Within each grain the current is homogeneous, and sharp changes are detected from grain to grain. As it can be observed, the highest currents are detected on grain B, indicating that it holds the thinnest *h*-BN stack on its surface. Grain C shows negligible currents similar to the electrical noise of the CAFM, indicating that the -2V applied were not enough to generate tunneling current across the *h*-BN stack. It should be emphasized that the current deviations within each grain are below one order of magnitude for all grains (compare maximum and minimum peaks within each grain in Figure 3c). This value is smaller than that observed in other thin insulating films (of similar thickness) being currently used in the industry, such as $HfO_2$ and $TiO_2$ (see Figure S3). Therefore, the electrical properties of h-BN within the same grain seem to be very homogeneous, which shows great potential to mitigate device-to-device variability problems of ultra scaled devices. Further electrical information about the grains has been obtained by measuring the onset voltage ($V_{ON}$) of the *h*-BN stacks on each Pt grain. The onset voltage is defined as the minimum voltage that needs to be applied between the CAFM tip and the substrate of the sample (Pt) in order to observe tunneling currents above the noise level.[27] Despite the noise level of our CAFM is ~ 1 pA, we selected $V_{ON} = V$ (I=5pA) in order to be completely sure that non-negligible current is flowing across the $h$-BN stack. For this experiment, $V_{ON}$ has been extracted by measuring individual current maps on each grain (1 μm × 1 μm). The results obtained (**Table S1**), strongly support the observations in Figure 3: the smallest $V_{ON}$ (0.1 V) was detected on grain B, and the highest (6 V) on grain C.

In Figure 3c it is difficult to quantify the real thickness of the h-BN on each Pt grain, and correlate it with the current levels observed. The reason is that best techniques used for physical thickness characterization of 2D materials (i.e. TEM) are destructive. In order to provide more insights to this point here we perform an additional analysis, consisting on measuring I-V curves at different locations of each grain via CAFM. Based on the shape of the I-V curves showing the tunneling current across the insulating stack, its physical thickness can be calculated very accurately (this was done before for ultra thin $SiO_2$ films with sub-nanometer resolution.[28] In previous works[29] it was determined that the dominant conduction across multilayer h-BN stacks was by Forler-Nordheim Tunneling (FNT). Therefore, here we use the FNT equation to calculate the tunneling current for different h-BN thicknesses:

$$I = \frac{A_{eff}\sqrt{m\phi_B}q^2V}{h^2d} \exp\left[\frac{-4\pi\sqrt{m\phi_B}d}{h}\right] \tag{1}$$

where $V$ is the applied voltage, $d$ is the thickness of the h-BN, $A_{eff}$ is the effective contact area, $\phi_B$ is the barrier height. The parameters $m$, $q$ and $h$ are the free electron mass, the electron charge and the Planck constant, respectively. This calculation has been repeated for different values of d depending on the number of layers (N), that is d = 0.33 nm for N=1, d = 0.66 nm for N=2, etc...

having increments of 0.33 nm for each layer until N=24. By comparing the calculations with the I-V curves experimentally collected (see Figure S4), it can be concluded that the tunneling currents across grain B fits well the conduction of monolayer h-BN, while the tunneling currents across grain D fit well the conduction across 11-13 layers (see Figure S4). This result is interesting because provides an indirect quantification of the thickness at each grain, something that the CAFM maps (Figure 3b) cannot provide.

Finally, the electrical properties of *h*-BN stacks grown on polycrystalline Pt substrates have been analyzed at the device level. To do so, squared Au/Ti top electrodes with lateral sizes ranging from 10 μm × 10 μm to 100 μm × 100 μm have been evaporated on the surface of the *h*-BN/Pt samples (see experimental section), leading to matrixes of metal/insulator/metal (MIM) cells (*i.e.* Au/Ti/*h*-BN/Pt). MIM cells are used in several devices, including FETs,[2] capacitors,[3] and memristors,[30] as well as in test structures to evaluate essential parameters of the insulator, such as charge trapping, stress induced leakage current, dielectric breakdown (BD) and resistive switching (RS).[31] Therefore, despite holding an easy structure, the MIM cells fabricated are representative of *h*-BN based microelectronic devices. Figure 4a shows the schematic of as-grown *h*-BN on polycrystalline Pt foil before and after electrodes deposition, in which *h*-BN film serves as insulating layer between the top (Au/Ti) and bottom (Pt) metal electrodes. The optical microscope image of the sample after Au/Ti electrode deposition is shown in Figure 4b. The corresponding EBSD map of the same area of the sample has been also collected (Figure 4c). It should be highlighted that, unlike in Figure 2b, in Figure 4c the EBSD map has been collected in the presence of *h*-BN on the Pt substrate. As the *h*-BN is an insulator, this may distort the signal related to the local crystallographic orientation. For this reason, the EBSD map in Figure 4d will be only used to distinguish different grains (in that case the contrast is large) but not to identify which is the real crystallographic orientation of the Pt within each grain. For this reason, the color scale in Figure 4c has been intentionally removed. Several devices on the same and different Pt grains have been tested in the probe station by applying ramped voltage stresses (RVS), and the resulting current vs. voltage (I-V) curves have been compared. As an example, Figure 4d shows the I-V curves collected on two devices with the same size (25 μm × 25 μm) that belong to the same Pt grain (devices 1 and 2 in Figure 4c). As it can be observed, the I-V characteristics for devices 1 and 2 are strikingly similar: *i)* the BD voltages ($V_{BD}$) show very small deviation (0.55 V and 0.61 V); *ii)* the pre- and post-BD currents ($I_{PRE-BD}$ and $I_{POST-BD}$, respectively) almost overlap, and *iii)* the $I_{POST-BD}/I_{PRE-BD}$ ratio is identical. These results are indeed indicating that the device-to-device variability within the same Pt grain is very small. Any potential variability of the pre-breakdown I-V curves in devices within the same Pt grain should be related to nanoscale inhomogeneities within the h-BN stack, such as thickness fluctuations (see Figure S2), local defects, h-BN domain boundaries and/or wrinkles. Nevertheless, as displayed in Figures 4c, 3c and S3 the variability of the electrical properties of the h-BN within the same Pt grain are very small. On the contrary, the devices with the same size but patterned on different Pt grains show very inhomogeneous I-V characteristics. As an example, Figure 4e shows the I-V curves collected on devices 3 and 4 (see Figure 4c). First, the pre-BD currents are very different; second, $V_{BD}$ for both devices are remarkably different: 2.09 V for device 3 and 0.89 V for device 4; and third, the $I_{POST-BD}/I_{PRE-BD}$ ratio is also slightly different. The different electrical properties of devices 3 and 4 are related to the different thicknesses of the *h*-BN stack, due to the different crystallographic orientation of the underlying Pt grain. These observations have been corroborated by measuring additional MIM devices at different Pt grains. As Figure 4f shows, the deviation of $V_{BD}$ within each grain is very small (from ± 0.013 V for grain C to

± 0.054 V for grain B), but the deviations of $V_{BD}$ from one Pt grain to another are large (from 0.58 V in grain A to 2.13 V in grain D). The breakdown event observed for all devices (displayed as a sharp current increase in Figures 4d and 4e) further demonstrates that the *h*-BN sheet grown on the Pt substrate is continuous, otherwise a larger current under smaller voltage (*i.e.* $10^{-2}$A @ 0.1V) typical of shorted devices should be observed.

Fortunately, cutting-edge electronic devices based on MIM cells cover ultra scaled areas. In the case of FETs, the current technology node is 7 nm, and the total length of current FETs never exceeds 50 nm. In the case of memristors and other non-volatile memories, such as resistive random access memories (RRAM), phase change random access memories (PCRAM) and ferroelectric random access memories (FeRAM), device areas down to 10 nm × 10 nm are preferred.[32] Therefore, as the diameter of the Pt grains easily surpass 100 μm, the fabrication of *h*-BN based electronic devices with very low variability is feasible. More efforts towards the growth of large-area single-crystalline *h*-BN stacks should conduct to ultra-low variability technologies.

In conclusion, the electrical homogeneity of *h*-BN stacks grown via CVD on Pt substrates has been analyzed by CAFM and a probe station. We observe that *h*-BN grows thicker on Pt grains with crystallographic orientations close to (101). *In situ* CAFM characterization reveals that the tunneling current across the *h*-BN grown on the same Pt grain is very homogeneous (*i.e.* more homogeneous than that observed in other insulators being currently used in the industry, such as $HfO_2$ and $TiO_2$), but sharp conductivity changes are detected from grain to grain. Device level tests in the probestation reveal that the variability of Au/Ti/*h*-BN/Pt devices within each grain is strikingly low in terms of $I_{PRE-BD}$, $I_{POST-BD}$ and $V_{BD}$. These results contribute to the understanding of the electrical properties of *h*-BN and variability of *h*-BN based electronic devices.

**Experimental Section**

*Growth of h-BN Pt substrates*: High purity (99.997%) 100 μm thick Pt foil purchased from Alfa Aesar (item no. 12059) was employed as substrate for the *h*-BN growth. First, the as-received Pt foil was cleaned in acetone and isopropanol (IPA) for 10 minutes to remove the surface contamination, and dried with a nitrogen gun. Then, the Pt foil was introduced in the center of the quartz tube, and the temperature ramped up to 950 °C in 70 sccm $H_2$ atmosphere. The time required to increase the temperature from ~20 °C to 950 °C was ~ 40 minutes, and following by the annealing process under the temperature of 950 °C in 70 sccm $H_2$ for 30 minutes, in order to remove the contamination contains carbon or oxygen. This pre-growth heating step is called annealing, and it is useful to clean impurities on the Pt surface. Then, the valve of the tube coming from the Borazine container (which used a $H_2$ flow rate of 0.1 sccm) was opened, allowing the introduction of borazine molecules in the quarz tube containing the Pt substrate. This process was kept for 1 hour at 950 °C. After that time, the temperature controller was set at room temperature and the CVD system was cooled down.

*Transfer process for the h-BN*: After sample fabrication, the *h*-BN stacks were transferred on $SiO_2$/Si wafers and TEM grids for analysis. To do so, liquid poly(methyl methacrylate) (PMMA) from MicroChem was spin-coated on both sides of the *h*-BN/Pt/*h*-BN sample at 2500 rpm for 1 min. The sample was backed in the oven at 70 °C for 10 min to solidify the PMMA, and the resulting sample (PMMA/*h*-BN/Pt/*h*-BN/PMMA) was immersed in a container filled with 1 M NaOH. In the same container, a Pt mesh was also introduced, and a potential difference of 3 V between it and the PMMA/*h*-BN/Pt/*h*-BN/PMMA sample was applied using a source meter. The Pt mesh served as anode, and the PMMA/*h*-BN/Pt/*h*-BN/PMMA sample as cathode. The application of voltage lead to the formation of hydrogen bubbles at the *h*-BN/Pt interface, leading to the effective PMMA/*h*-BN detachment from the Pt foil [33] in less than 10 minutes. Afterwards, the PMMA/*h*-BN stack was cleaned in deionized water three times to remove the residual NaOH solution. Finally, PMMA/*h*-BN was transferred on the target substrates (SiO$_2$/Si or TEM grids). When transferred on the SiO$_2$/Si substrate the PMMA was removed by soaking the entire sample in acetone for 2 hours, followed by an annealing at 400 °C for 2 h in a mixed H$_2$ (200 sccm) and Ar (200 sccm) atmosphere. When transferred on the TEM grids, only the annealing step was used (no sample soaking because that could damage the *h*-BN suspended on the perforated TEM grid).

*h-BN and Pt characterization*: The different Pt grains and their crystallographic orientations were analyzed using a standard optical microscope, and an EBSD system (from Oxford Technology) integrated in a scanning electron microscope (Zeiss Merlin HRSEM). The topographic maps in Figures S1-S2 were collected using a Dimension 3000 AFM working in air atmosphere. These experiments were performed in tapping mode using Si probe tips from Budgetsensors (model: Tap300-G). The CAFM experiments were carried out in a Multimode VI equipment from Veeco working inside a nitrogen chamber (relative humidity ~ 0.5%).[34] The use of a nitrogen atmosphere is beneficial to stabilize the current signal and achieve larger lateral resolution.[35] For this experiment we used PtIr varnished Si probes from Nanosensors (model: CONTPT). The cross section in the current CAFM maps have been calculated using the NanoScope Analysis software of the AFM (Bruker) and assembled using OriginPro 8 software. Atomic scale information about the thickness and quality of the *h*-BN stacks was collected using a JEOL 2010 HRTEM with EELS capability integrated, and a LabRAM Raman spectrometer from Horiba.

*Device fabrication and characterization*: Squared metallic top electrodes with lateral sizes ranging from 10 μm × 10 μm to 100 μm × 100 μm have been deposited on as-grown *h*-BN/Pt samples. First 20 nm Ti and second 60 nm Au have been deposited via E-beam evaporator (Ajaint AJA-ATC) using a laser-patterned shadow mask. The electrical measurements were carried out in a Summit 11000 AP probe station connected to an Agilent 4155C semiconductor parameter analyzer. The RVS was applied to the top electrodes and the Pt substrates were grounded.

**Supporting Information**.

The Supporting Information are available free of charge.

Additional AFM characterization of the *h*-BN/300-nm-$SiO_2$/Si sample and Electrical measurements are conducted on each grain will be found in Supporting Information. (PDF)

AUTHOR INFORMATION

**Corresponding Author**

Mario Lanza, email: mlanza@suda.edu.cn

**Author Contributions**

The manuscript was written through contributions of all authors. All authors have given approval to the final version of the manuscript. ‡These authors contributed equally.

## ACKNOWLEDGMENT

Fei Hui and Wenjing Fang contributed equally to this work. F. Hui acknowledges the support from the Young 1000 Global Talent Recruitment Program of the Ministry of Education of China, the National Natural Science Foundation of China (grants no. 61502326, 41550110223, 11661131002), the Jiangsu Government (grant no. BK20150343), the Ministry of Finance of China (grant no. SX21400213) and the Young 973 National Program of the Chinese Ministry of Science and Technology (grant no. 2015CB932700). The Collaborative Innovation Center of Suzhou Nano Science & Technology, the Jiangsu Key Laboratory for Carbon-Based Functional Materials & Devices, the Priority Academic Program Development of Jiangsu Higher Education Institutions, and the Opening Project of Key Laboratory of Microelectronic Devices & Integrated Technology (Institute of Microelectronics, Chinese Academy of Sciences) are also acknowledged. W. Fang, T. Kpulun, G. Harris and J. Kong acknowledge the support from the STC Center for Integrated Quantum Materials, NSF Grant No. DMR-1231319. H. Wang and J. Kong acknowledge the support from NSF DMR/ECCS – 1509197. W.S. Leong acknowledges the support from SUTD-MIT Postdoctoral Fellows Program.

FIGURES

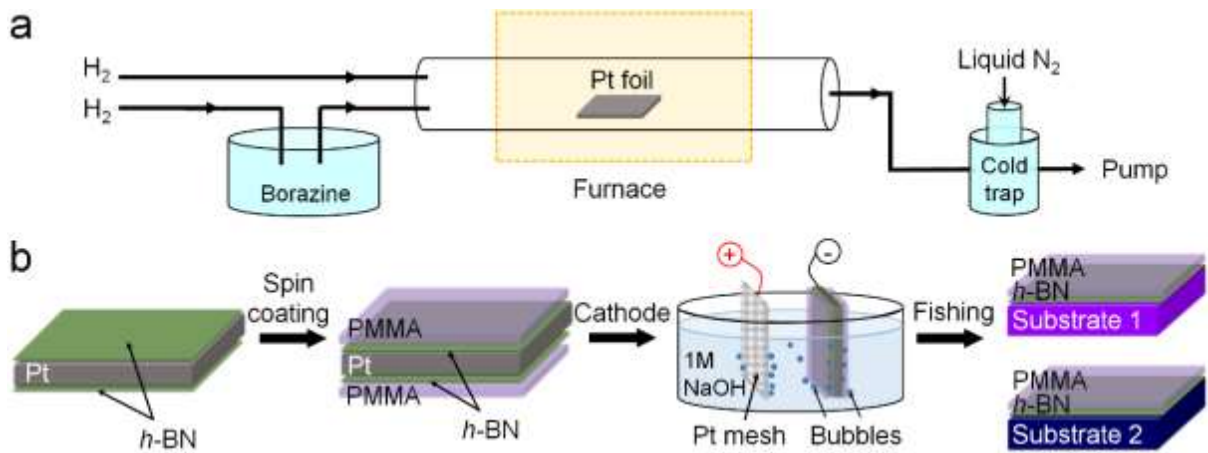

**Figure 1.** (a) Schematic of LPCVD process to grow *h*-BN on Pt foils. (b) Electrochemical (bubble) method to transfer *h*-BN from Pt foil to the target substrates.

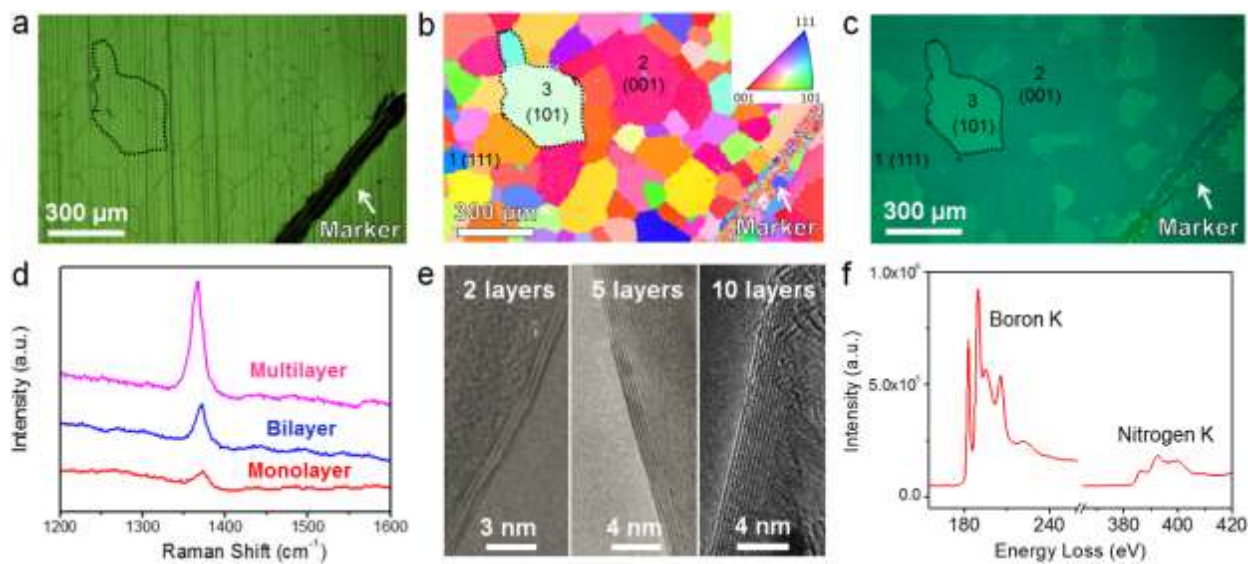

**Figure 2.** Characterization of *h*-BN films. (a) Optical microscope and (b) EBSD images of a polycrystalline Pt substrate before *h*-BN growth (annealed). The two images correspond to the same location of the sample, as highlighted by the marker in the right-bottom part. (c) Optical microscope image of a *h*-BN/300-nm-SiO₂/Si sample. The *h*-BN was grown on the area displayed in (a). The granular pattern can be distinguished with different dark-green and light-green colors. (d) Raman spectra collected on different locations of the *h*-BN/300-nm-SiO₂/Si sample. Monolayer/multilayer have been collected at dark-green/light-green locations of panel (c). High resolution TEM images demonstrating the different thicknesses and the good layered structure of the *h*-BN stacks. (f) EELS spectrum of *h*-BN showing the typical boron and nitrogen peaks.

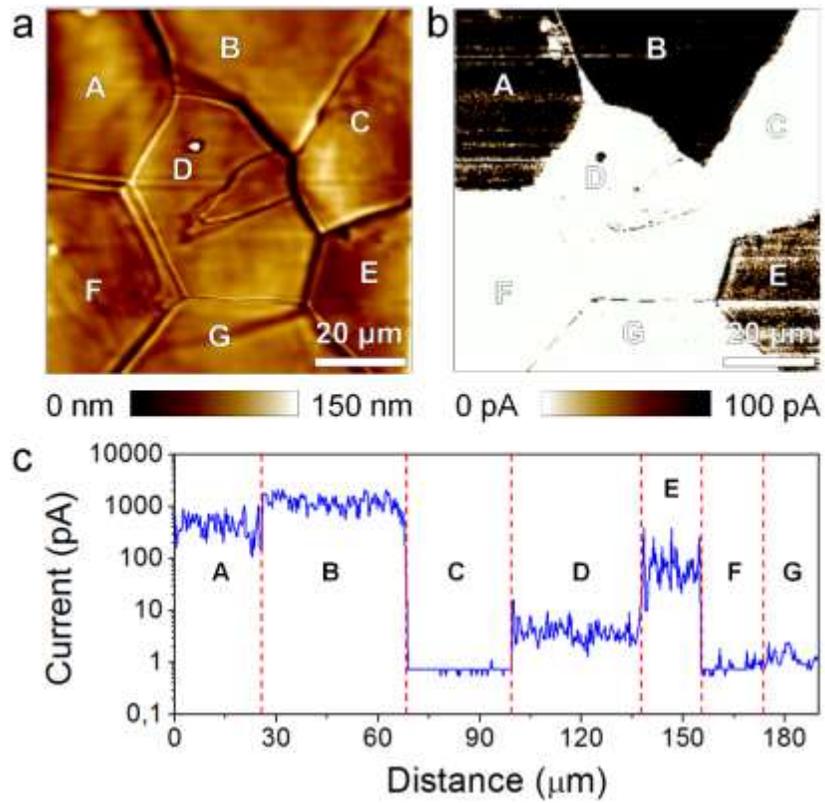

**Figure 3.** CAFM Characterization of as-grown *h*-BN/Pt. (a) Topographic and (b) current maps simultaneously collected on *h*-BN grown on a polycrystalline Pt substrate, under a bias of -2V (applied to the substrate, tip grounded). (c) Assembly of cross sections collected at the different grains of the current map in (b).

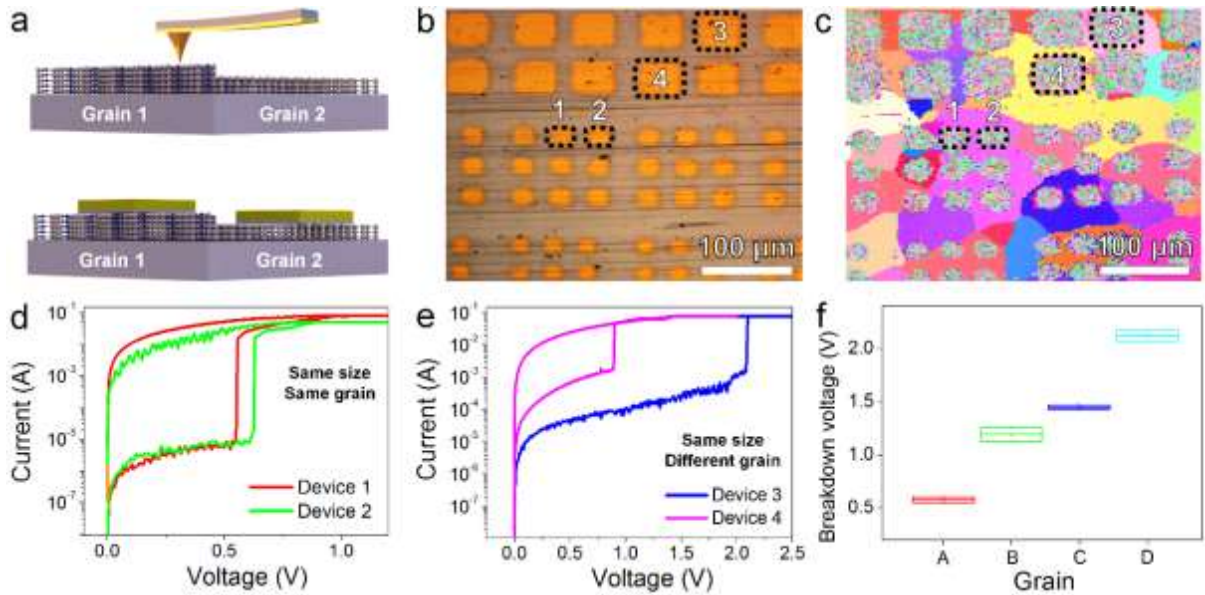

**Figure 4.** Device level characterization of *h*-BN stacks. (a) Schematic of as-grown *h*-BN/Pt (top) and *h*-BN based devices (bottom). The device cell is an Au/Ti/*h*-BN/Pt structure. (b) Optical image and (c) EBSD map of matrixes of Au/Ti/*h*-BN/Pt cells with different lateral sizes. As this map has been collected with the presence of *h*-BN, it is only meaningful to distinguish different grains, not to assess the real crystallographic orientation of each grain. For this reason, the color scale has been intentionally removed. (d) I-V curves collected on two devices with the same size (25 μm × 25 μm) within the same Pt grain. (e) I-V curves collected on two devices with the same size (50 μm × 50 μm) located at different Pt grains. (f) Analysis of the BD voltage for Au/Ti/*h*-BN/Pt devices patterned on 4 different Pt grains (located outside the region of 4c).

# Electrical homogeneity of chemical vapor deposited hexagonal boron nitride

*Fei Hui,\*,§,‡ Wenjing Fang,§,‡ Wei Sun Leong, §,& Tewa Kpulun,ξ Haozhe Wang,§ Hui Ying*

*Yang,& Marco A. Villena\*, Gary Harris,ξ Jing Kong,§ Mario Lanza\**

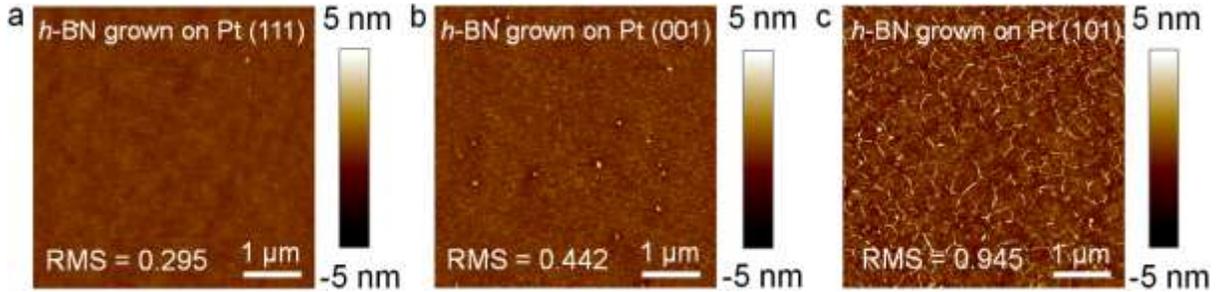

**Figure S1**. AFM characterization of the *h*-BN/300-nm-SiO₂/Si sample displayed in Figure 2c. The *h*-BN has been grown on via CVD on a polycrystalline Pt substrate and transferred on a 300-nm-SiO₂/Si sample. After the transfer, the optical microscope photograph (Figure 2c) shows shapes perfectly matching the Pt grains observed via EBSD (Figure 2b). Therefore, it is possible to know the crystallographic orientation of the Pt substrate on which the *h*-BN was grown. Panels (a)-(c) in Figure S1 show that the roughness and density of wrinkles in the *h*-BN strongly depend on the crystallographic orientation of the Pt substrate on which it was grown.

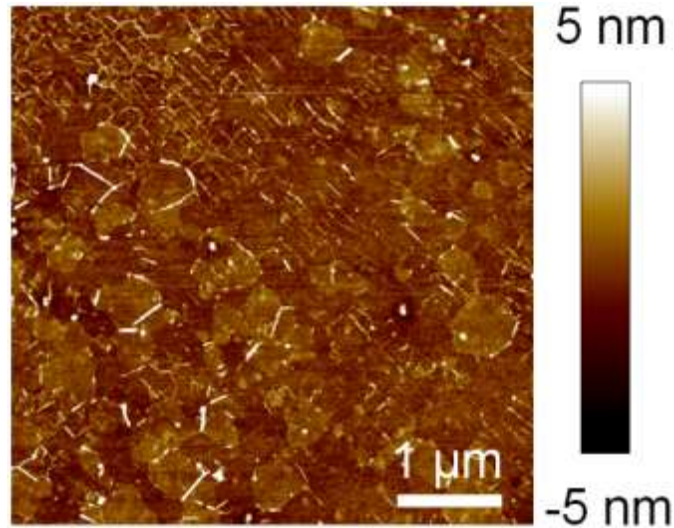

**Figure S2**. AFM characterization of a *h*-BN/300-nm-SiO₂/Si sample, on a region on which the h-BN was previously grown on Pt(101). This is  Zoom-in image of Figure S1c. This image shows multilayer *h*-BN islands.

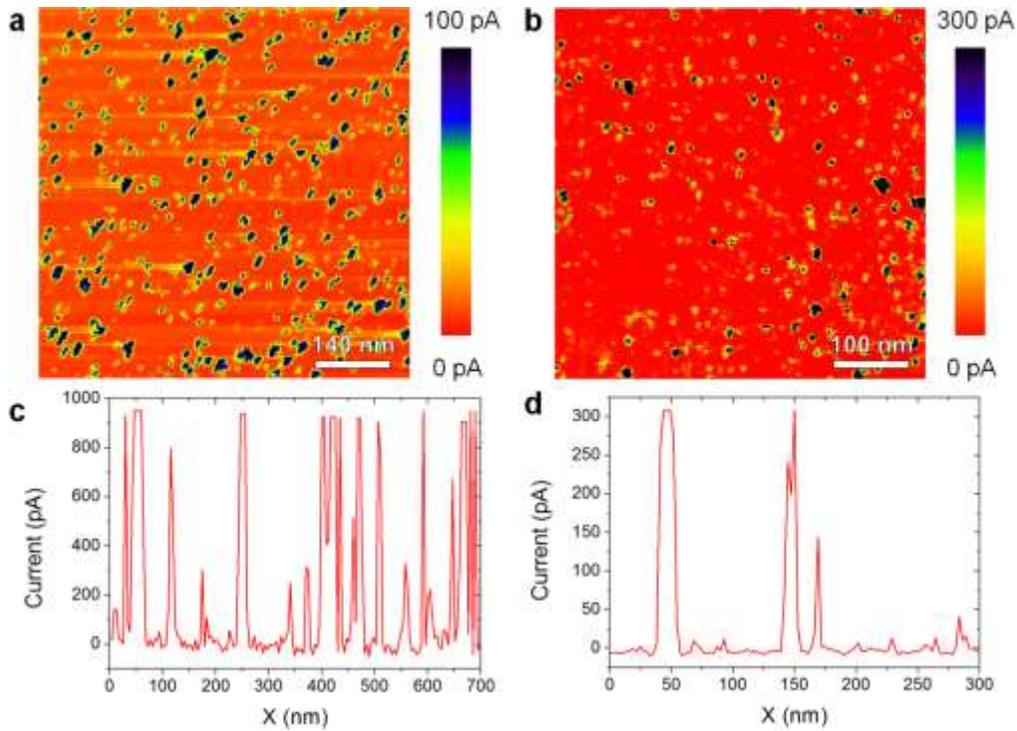

**Figure S3.** Current maps collected with the tip of the CAFM on the surface of (a) 4 nm thick HfO$_2$ and (b) 2 nm thick TiO$_2$ films, both of them grown via atomic layer deposition on a conductive substrate (Zr below the HfO$_2$ and n++Si below the TiO$_2$). The current fluctuations are 2-3 orders of magnitude. (c) and (d) show the cross sections of the maps in (a) and (b) respectively.

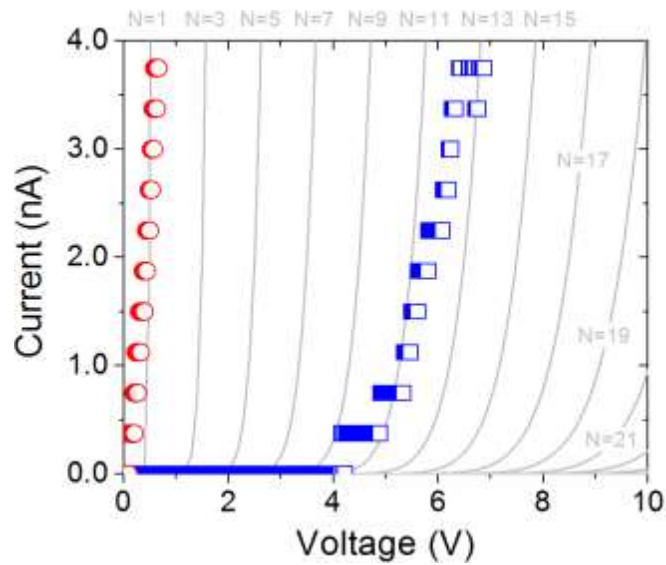

**Figure S4.** Calculation of FNT conduction for different thicknesses (N is the number of layers). Typical experimental IV curves measured on grain B (red symbols) and grain D (blue symbols) of Figure 3b of the manuscript.

| Grain | Current (for a voltage of -1 V) | Voltage (for a current of 5 pA) |
|:---:|:---:|:---:|
| A | 5 nA | 0.5 V |
| B | 5 nA | 0.1 V |
| C | < 1pA | 6 V |
| D | < 1pA | 2 V |
| E | < 1pA | 1 V |
| F | < 1pA | 5 V |
| G | < 1pA | 5 V |

**Table S1** Electrical measurements are conducted on each grain with the constant voltage and constant 5 pA, respectively.